# The Linear Parameters and the Decoupling Matrix for Linearly Coupled Motion in 6 Dimensional Phase Space

George Parzen

February 17, 1995

It will be shown that starting from a coordinate system where the 6 phase space coordinates are linearly coupled, one can go to a new coordinate system where the motion is uncoupled by means of a linear transformation. The original coupled coordinates and the new uncoupled coordinates are related by a 6x6 matrix, R. R will be called the decoupling matrix. It will be shown that of the 36 elements of the 6x6 decoupling matrix R, only 12 elements are independent. This may be contrasted with the results for motion in 4-dimensional phase space, where R has 4 independent elements. A set of equations is given from which the 12 elements of R can be computed from the one period transfer matrix. This set of equations also allows the linear parameters for the uncoupled coordinates to be computed from the one period transfer matrix. An alternative procedure for computing the linear parameters and the 12 independent elements of the decoupling matrix R is also given which depends on computing the eigenvectors of the one period transfer matrix. These results can be used in a tracking program, where the one period transfer matrix can be computed by multiplying the transfer matrices of all the elements in a period, to compute the linear parameters and the elements of the decoupling matrix R.

---

# 1. Introduction

It will be shown that starting from a coordinate system where the 6 phase space coordinates are linearly coupled, one can go to a new coordinate system, where the motion is uncoupled, by means of a linear transformation. The original coupled coordinates and the new uncoupled coordinates are related by a 6 × 6 matrix, $R$. $R$ will be called the decoupling matrix.

It will be shown that of the 36 elements of the 6 × 6 decoupling matrix $R$, only 12 elements are independent. This may be contrasted with the results[1] for motion in 4-dimensional phase space, where $R$ has 4 independent elements. A set of equations is given from which the 12 elements of $R$ can be computed from the one period transfer matrix. This set of equations also allows the linear parameters, the $\beta_i, \alpha_i, i = 1, 3$, for the uncoupled coordinates, to be computed from the one period transfer matrix.

An alternative procedure for computing the linear parameters, $\beta_i, \alpha_i, i = 1, 3$, and the 12 independent elements of the decoupling matrix $R$ is also given which depends on computing the eigenvectors of the one period transfer matrix.

These results can be used in a tracking program, where the one period transfer matrix can be computed by multiplying the transfer matrices of all the elements in a period, to compute the linear parameters $\alpha_i$ and $\beta_i$, $i = 1, 3$, and the elements of the decoupling matrix $R$.

The procedure presented here for studying coupled motion in 6-dimensional phase space can also be applied to coupled motion in 4-dimensional phase space, where it may be a useful alternative procedure to the procedure presented by Edwards and Teng[1]. In particular, it gives a simpler programing procedure for computing the beta functions and the emittances for coupled motion in 4-dimensional phase space.

## 2. The Decoupling Matrix, $R$

The particle coordinates are assumed to be $x$, $p_x$, $y$, $p_y$, $z$, $p_z$. The particle is acted upon by periodic fields that couple the 6 coordinates. The linearized equations of motion are assumed to be

$$\frac{dx}{ds} = A(s)\,x$$

$$x = \begin{bmatrix} x \\ p_x \\ y \\ p_y \\ z \\ p_z \end{bmatrix}, \qquad (2\text{-}1)$$

where the $6 \times 6$ matrix $A(s)$ is assumed to be periodic in $s$ with the period $L$. Note that the symbol $x$ is used to indicate both the column vector $x$ and the first element of this column vector. The meaning of $x$ should be clear from the context. This kind of double use of a symbol will be used in several places in this paper. The $6 \times 6$ transfer matrix $\mathrm{T}(s, s_0)$ obeys

$$x(s) = \mathrm{T}(s, s_0)\,x(s_0)$$
$$\frac{d\mathrm{T}}{ds} = A(s)\,\mathrm{T} \qquad (2\text{-}2)$$

It is assumed that the motion is symplectic so that

$$\mathrm{T}\overline{\mathrm{T}} = I$$
$$\overline{\mathrm{T}} = \widetilde{\mathrm{S}}\widetilde{\mathrm{T}}\,\mathrm{S} \qquad (2\text{-}3)$$

where $I$ is the $6 \times 6$ identity matrix, $\widetilde{\mathrm{T}}$ is the transpose of T and the $6 \times 6$ matrix S is given by

$$\mathrm{S} = \begin{bmatrix} 0 & 1 & 0 & 0 & 0 & 0 \\ -1 & 0 & 0 & 0 & 0 & 0 \\ 0 & 0 & 0 & 1 & 0 & 0 \\ 0 & 0 & -1 & 0 & 0 & 0 \\ 0 & 0 & 0 & 0 & 0 & 1 \\ 0 & 0 & 0 & 0 & -1 & 0 \end{bmatrix} \qquad (2\text{-}4a)$$

Note that

$$\widetilde{\mathrm{S}} = -\mathrm{S}, \quad \widetilde{\mathrm{S}} = \mathrm{S}^{-1}, \quad \mathrm{S}^2 = -1 \qquad (2\text{-}4b)$$

and S can be written in terms of $2 \times 2$ matrices as

$$\mathrm{S} = \begin{bmatrix} \mathrm{S} & 0 & 0 \\ 0 & \mathrm{S} & 0 \\ 0 & 0 & \mathrm{S} \end{bmatrix} \qquad (2\text{-}4c)$$

where the 2 × 2 matrix S is given by

$$S = \begin{bmatrix} 0 & 1 \\ -1 & 0 \end{bmatrix} \tag{2-4d}$$

Note that the same symbol S is used to indicate the 6 × 6 and 2 × 2 symplectic matrix. The meaning of S should be clear from the context.

The 6×6 transfer matrix $T(s, s_0)$ has 36 elements. However, the number of independent elements is smaller because of the symplectic conditions given by Eq. (2-3). There are 15 symplectic conditions or $(k^2 - k)/2$ where $k = 6$. This can be seen by noting that the equation $T\overline{T} = I$ can be written as

$$\begin{aligned} D &= T\,\tilde{S}\tilde{T} - S = 0 \\ \tilde{D} &= -D \end{aligned} \tag{2-5}$$

Since $D$ is anti-symmetric, it has $(36 - 6)/2 = 15$ independent elements. Eq. (2-5) then represents 15 symplectic conditions. The transfer matrix T then has 21 independent elements.

One can also introduce the one period transfer matrix $\hat{T}(s)$ defined by

$$\hat{T}(s) = T(s + L, s) \tag{2-6}$$

$\hat{T}(s)$ is also symplectic and has 21 independent elements.

The 15 symplectic conditions for T can be written down more explicitly. T can be written in terms of the 2 × 2 matrices $T_{ij}, i = 1, 3, j = 1, 3$ as

$$T = \begin{bmatrix} T_{11} & T_{12} & T_{13} \\ T_{21} & T_{22} & T_{23} \\ T_{31} & T_{32} & T_{33} \end{bmatrix} \tag{2-7a}$$

Using $\overline{T} = \tilde{S}\tilde{T}S$, one can show that $\overline{T}$ is given by

$$\overline{T} = \begin{bmatrix} \overline{T}_{11} & \overline{T}_{21} & \overline{T}_{31} \\ \overline{T}_{12} & \overline{T}_{22} & \overline{T}_{32} \\ \overline{T}_{13} & \overline{T}_{23} & \overline{T}_{33} \end{bmatrix} \tag{2-7b}$$

Then $T\overline{T} = I$ gives the conditions

$$\sum_{k=1,3} T_{ik}\overline{T}_{jk} = \delta_{ij} \qquad i = 1, 3 \quad j = 1, 3 \tag{2-8a}$$

which can be written as

$$|T_{11}| + |T_{12}| + |T_{13}| = 1$$
$$|T_{21}| + |T_{22}| + |T_{23}| = 1$$
$$|T_{31}| + |T_{32}| + |T_{33}| = 1$$
$$T_{11}\overline{T}_{21} + T_{12}\overline{T}_{22} + T_{13}\overline{T}_{23} = 0$$
$$T_{11}\overline{T}_{31} + T_{12}\overline{T}_{32} + T_{13}\overline{T}_{33} = 0$$
$$T_{21}\overline{T}_{31} + T_{22}\overline{T}_{32} + T_{23}\overline{T}_{33} = 0$$

(2-8b)

To get the first 3 equations, one uses the result for 2 × 2 matrices

$$T_{ij}\overline{T}_{ij} = |T_{ij}|$$

where $|T_{ij}|$ is the determinant of $T_{ij}$. Eqs. (2.8) are the 15 symplectic conditions for T.

One now goes to a new coordinate system where the particle motion is not coupled. The coordinates in the uncoupled coordinate system will be labeled $u, p_u, v, p_v, w, p_w$. It is assumed that the original coupled coordinate system and the new uncoupled coordinate system are related by a 6 × 6 matrix $R(s)$

$$x = R\,u$$

$$u = \begin{bmatrix} u \\ p_u \\ v \\ p_v \\ w \\ p_w \end{bmatrix}$$

(2-9)

$R(s)$ will be called the decoupling matrix.

One can introduce a 6 × 6 transfer matrix for the uncoupled coordinates called $P(s, s_0)$ such that

$$u(s) = P(s, s_0)\,u \tag{2-10a}$$

and one finds that

$$P(s, s_0) = R^{-1}(s)\,T(s, s_0)\,R(s_0) \tag{2-10b}$$

one can also introduce the one period transfer matrix $\hat{P}(s)$ defined by

$$\hat{P}(s) = P(s + L, s)$$
$$\hat{P}(s) = R^{-1}(s + L)\,\hat{T}(s)\,R(s) \tag{2-11}$$

The decoupling matrix is defined as the $6 \times 6$ matrix that diagonalize $\hat{P}(s)$, which means here that when the $6 \times 6$ matrix $\hat{P}$ is written in terms of $2 \times 2$ matrices it has the form

$$\hat{P} = \begin{bmatrix} \hat{P}_{11} & 0 & 0 \\ 0 & \hat{P}_{22} & 0 \\ 0 & 0 & \hat{P}_{33} \end{bmatrix} \tag{2-12}$$

where $\hat{P}_{ij}$ are $2 \times 2$ matrices. $\hat{P}$ will be called a diagonal matrix in the sense of Eq. (2-12).

The definition given so far of the decoupling matrix $R$, will be seen to not uniquely define $R$ and one can add the two conditions on $R$ that it is a symplectic matrix and it is a periodic matrix. Thus

$$R\,\overline{R} = I$$
$$\overline{R} = \tilde{S}\,\tilde{R}\,S \tag{2-13}$$
$$R(s+L) = R(s)$$

The justification for the above is given by the solution found for $R(s)$ below in section 4.

Because $T(s, s_0)$ and $R(s)$ are symplectic, it follows that $P(s, s_0)$ and $\hat{P}(s)$ are symplectic. Eq. (2-8) and (2-9) can be rewritten as

$$P(s, s_0) = \overline{R}(s)\,T(s, s_0)\,R(s_0) \tag{2-14a}$$
$$\hat{P}(s) = \overline{R}(s)\,\hat{T}(s)\,R(s) \tag{2-14b}$$

It also follows that the $2 \times 2$ matrices has 3 independent elements as $|\hat{P}_{11}| = |\hat{P}_{22}| = |\hat{P}_{33}| = 1$. Eq. (2-14b) can be written as

$$\hat{T}(s) = R(s)\,\hat{P}(s)\,\overline{R}(s) \tag{2-15}$$

Eq. (2-15) shows that $R$ has 12 independent elements, as $\hat{T}$ has 21 independent elements and $\hat{P}$ has 9 independent elements. As $R$ has only 12 independent elements, one can suggest that $R$ has the form, when written in terms of $2 \times 2$ matrices,

$$R = \begin{bmatrix} q_1 I & R_{12} & R_{13} \\ R_{21} & q_2 I & R_{23} \\ R_{31} & R_{32} & q_3 I \end{bmatrix} \tag{2-16}$$

where $q_1, q_2, q_3$ are scalar quantities, the $R_{ij}$ are $2 \times 2$ matrices and $I$ is the $2 \times 2$ identity matrix. The matrix in Eq. (2-16) appears to have 27 independent elements. However, $R$ is symplectic and has to obey the 15 symplectic conditions, and this reduces the number of independent elements to 12. The justification for assuming this form of $R$, given by Eq. (2-16), will be provided below where a solution for $R$ will be found assuming this form for $R$.

The 15 symplectic conditions for $R$ may be written, using Eq. (2-8b), as

$$\begin{aligned}
q_1^2 + |R_{12}| + |R_{13}| &= 1 \\
|R_{21}| + q_2^2 + |R_{23}| &= 1 \\
|R_{31}| + |R_{32}| + q_3^2 &= 1 \\
q_1 \overline{R}_{21} + R_{12} q_2 + R_{13} \overline{R}_{23} &= 0 \\
q_1 \overline{R}_{31} + R_{12} \overline{R}_{32} + R_{13} q_3 &= 0 \\
R_{21} \overline{R}_{31} + q_2 \overline{R}_{32} + R_{23} q_3 &= 0
\end{aligned} \quad (2\text{-}17)$$

Using Eq. (2-16) for $R$ and the symplectic conditions Eq. (2-17), one can, in principle, solve Eq. (2-15) for $R$ and $\hat{P}$ in terms of the one period matrix $\hat{T}$. This was done by Edwards and Teng[1] for motion in 4-dimensional phase space where $\hat{T}$ has 10 independent elements, $R$ has 4 independent elements and $\hat{P}$ has 6 independent elements. An analytical solution of Eq. (2-15) for the 6-dimensional case was not found. However, a different procedure for finding $\hat{P}$ and $R$ will be given below in section 4 by finding the eigenvectors of $\hat{P}$, using the eigenvectors of the one period matrix, $\hat{T}$.

The $2 \times 2$ matrices $P_{11}$, $P_{22}$, $P_{33}$ which make up $\hat{P}$ each have 3 independent elements and can be written in the form

$$\hat{P}_{11} = \begin{bmatrix} \cos \psi_1 + \alpha_1 \sin \psi_1 & \beta_1 \sin \psi_1 \\ -1/\gamma_1 \sin \psi_1 & \cos \psi_1 - \alpha_1 \sin \psi_1 \end{bmatrix} \quad (2\text{-}18)$$
$$\gamma_1 = \left(1 + \alpha_1^2\right)/\beta_1$$

with similar expressions for $\hat{P}_{22}$ and $\hat{P}_{33}$ Eq. (2-18) and the similar expressions for $\hat{P}_{22}$, $\hat{P}_{33}$ can be used to define the three beta functions $\beta_1$, $\beta_2$ and $\beta_3$. This definition of the beta functions will be further justified below where one finds expressions for the three emittance invariants $\epsilon_1$, $\epsilon_2$ and $\epsilon_3$ in terms of $\beta_1$, $\alpha_1$, $\beta_2$, $\alpha_2$ and $\beta_3$, $\alpha_3$. If one could solve Eq. (2-15) for $R$ and $\hat{P}$ in terms of the one period transfer matrix $\hat{T}$, then Eq. (2-18) could be used to find the beta functions $\beta_i$ from the one period transfer matrix. A different procedure for finding the linear parameters $\beta_1$, $\alpha_1$, $\beta_2$, $\alpha_2$ and $\beta_3$, $\alpha_3$ will be given below.

## 3. The Linear Parameters $\beta$, $\alpha$, and $\psi$ and the Eigenvectors of the Transfer Matrix

In this section, the eigenvectors of the one period transfer matrix, $\hat{P}$, will be found and expressed in terms of the linear periodic parameters $\beta$, $\alpha$ and $\psi$. These will be used below to compute the linear parameters from the one period transfer matrix $\hat{T}$. They will also be used to find the three emittance invariants $\epsilon_1$, $\epsilon_2$ and $\epsilon_3$ and to express them in terms of the linear parameters $\beta_i$, $\alpha_i$, $i = 1, 3$.

The uncoupled transfer matrix obeys

$$\frac{d}{ds} = P(s, s_0) = B(s)\ P(s, s_0)$$
$$B = \overline{R}\ A\ R + \frac{d\overline{R}}{ds} R \tag{3-1}$$

This follows from Eq. (2-2) and Eq. (2-14).

One sees from Eq. (3-1) that $B(s)$ is a periodic matrix, $B(s + L) = B(s)$. It can also be shown that $B$ is a diagonal matrix in the sense that written in terms of $2 \times 2$ matrices, it has the form

$$B = \begin{bmatrix} B_{11} & 0 & 0 \\ 0 & B_{22} & 0 \\ 0 & 0 & B_{33} \end{bmatrix} \tag{3-2}$$

To establish Eq. (3-2), one needs the equation for $d\hat{P}/ds$

$$\frac{d\hat{P}}{ds} = B\ \hat{P} - \hat{P}\ B \tag{3-3}$$

Eq. (3-3) follows from the result

$$\hat{P}(s) = P(s, s_0)\ \hat{P}(s_0)\ P(s_0, s) \tag{3-4}$$

which only requires that $P(s + L, s_0 + L) = P(s, s_0)$, which follows from $P(s, s_0) = R(s)T(s, s_0)\overline{R}(s_0)$ and $T(s + L, s_0 + L) = T(s, s_0)$. Eq. (3-4) when differentiated gives Eq. (3-3), using the results

$$\frac{d}{ds}P(s, s_0) = B(s) P(s, s_0)$$
$$\frac{d}{ds}P(s_0, s) = -P(s_0, s) B(s) \tag{3-5}$$

The last equation follows from $P(s, s_0)\ P(s_0, s) = I$. One can show that $B_{12} = 0$ by using Eq. (3-3)

$$\frac{d}{ds}\hat{P}_{12} = 0 = B_{12}\hat{P}_{22} - \hat{P}_{11}B_{12} = B_{12}\left(\hat{P}_{22} - \hat{P}_{11}\right)$$
$$B_{12} = 0 \tag{3-6}$$

In the same way one can show that $B_{ij} = 0$ if $i \neq j$. It has now been established that the uncoupled coordinates indicated by the column vector $u$ obeys the equations of motion

$$\frac{du}{ds} = B\, u \qquad (3\text{-}7a)$$

which can be written as

$$\frac{d}{ds}\begin{bmatrix}\overline{u}\\\overline{v}\\\overline{w}\end{bmatrix} = \begin{bmatrix}B_{11} & 0 & 0\\ 0 & B_{22} & 0\\ 0 & 0 & B_{33}\end{bmatrix}\begin{bmatrix}\overline{u}\\\overline{v}\\\overline{w}\end{bmatrix}$$
$$\overline{u} = \begin{bmatrix}u\\p_u\end{bmatrix}, \overline{v} = \begin{bmatrix}v\\p_v\end{bmatrix}, \overline{w} = \begin{bmatrix}w\\p_w\end{bmatrix} \qquad (3\text{-}7b)$$

The three sets of coordinates, $\overline{u}, \overline{v}, \overline{w}$ are uncoupled, and the $\overline{u}$ coordinates obeys

$$\frac{d}{ds}\overline{u} = B_{11}\overline{u} \qquad (3\text{-}8)$$

As the $2 \times 2$ matrix $B_{11}$ is periodic, one can show[2] that the eigenvector of the transfer matrix for $\overline{u}$ is

$$\overline{u}_1 = \begin{bmatrix}\beta_1^{1/2}\\ \beta_1^{1/2}(-\alpha_1 + i)\end{bmatrix}\exp(i\psi_1) \qquad (3\text{-}9a)$$
$$\widetilde{\overline{u}}_1^* \, S \, u_1 = 2i$$

with the eigenvalue $\lambda_1 = \exp(i\mu_1)$. $\beta_1(s)$, $\alpha_1(s)$ are periodic functions and the phase function $\psi_1 = \mu_1 s/L + g_1(s)$ where $g_1(s)$ is periodic. One may notice that the $B_{11}$ matrix does not have the properties usually assumed in the large accelerator approximation, that $(B_{11})_{11} = (B_{21})_{22} = 0$ and $(B_{11})_{12} = 1$. Thus obtaining Eq. (3-9) requires a study[2] of the more general equation (3-8). In particular the relationship between $\beta_1$, $\alpha_1$, $\psi_1$ are now given by[2]

$$\psi_1 = \int_0^s (B_{11})_{12}\frac{ds}{\beta_1}$$
$$\alpha_1 = \frac{1}{(B_{11})_{12}}\left(-\frac{1}{2}\frac{d\beta_1}{ds} + (B_{11})_{11}\beta_1\right) \qquad (3\text{-}9b)$$

One can write down similar expressions for the eigenfunctions of the transfer matrix of $\overline{v}$ and $\overline{w}$, in terms of $\beta_2$, $\alpha_2$, $\psi_2$ and $\beta_3$, $\alpha_3$, $\psi_3$, with the eigenvalues $\lambda_2 = \exp(i\mu_2)$, $\lambda_3 = \exp(i\mu_3)$.

One can now write down the eigenvectors of the $\hat{P}$ matrix using Eq. (3-9). These eigenvectors will be called $u_1, u_2, u_3, u_4, u_5, u_6$, each of which is a $6 \times 1$ column vector.

$$u_1 = \begin{bmatrix} \beta_1^{1/2} \\ \beta_1^{-1/2}(-\alpha_1 + i) \\ 0 \\ 0 \\ 0 \\ 0 \end{bmatrix} \exp(i\psi_1), \quad u_3 = \begin{bmatrix} 0 \\ 0 \\ \beta_2^{1/2} \\ \beta_2^{-1/2}(-\alpha_2 + i) \\ 0 \\ 0 \end{bmatrix} \exp(i\psi_2),$$

$$u_3 = \begin{bmatrix} 0 \\ 0 \\ 0 \\ 0 \\ \beta_3^{1/2} \\ \beta_3^{-1/2}(-\alpha_3 + i) \end{bmatrix} \exp(i\psi_3)$$

(3-10)

$$u_2 = u_1^*, \quad u_4 = u_3^*, \quad u_6 = u_5^*$$

$$\widetilde{u}_1^* S u_1 = \widetilde{u}_3^* S u_3 = \widetilde{u}_5^* S u_5 = 2i$$

with the eigenvalues $\lambda_1 = \exp(i\mu_1)$, $\lambda_3 = \exp(i\mu_2)$, $\lambda_5 = \exp(i\mu_3)$, $\lambda_2 = \lambda_1^*$, $\lambda_4 = \lambda_3^*$ and $\lambda_6 = \lambda_5^*$.

These results for the eigenvectors will be used below to express the 3 emittance invariants $\epsilon_1, \epsilon_2, \epsilon_3$, in terms of the linear parameters $\beta_i, \alpha_i, \psi_i, i = 1, 3$, and to find a procedure for finding the linear parameters $\beta_i, \alpha_i, \psi_i$ from the one period transfer matrix $\hat{T}$.

## 4. Computing the Linear Parameters $\beta$, $\alpha$, $\psi$ from the Transfer Matrix

An important problem in tracking studies is how to compute the linear parameters, $\beta$, $\alpha$, $\psi$, defined in section 3, from the one period transfer matrix. The one period transfer matrix can be found by multiplying the transfer matrices of each of the elements in a period. A procedure is given below for computing the linear parameters, which also computes the decoupling matrix $R$ from the one period transfer matrix.

The first step in this procedure is to compute the eigenvectors and their corresponding eigenvalues for the one period transfer matrix $\hat{T}$. This can be done using one of the standard routines available for finding the eigenvectors of a real matrix. $\hat{T}$ is assumed to be known. In this case, there are 6 eigenvectors indicated by the 6 column vectors $x_1, x_2, x_3, x_4, x_5$ and $x_6$. Because $\hat{T}$ is a real $6 \times 6$ matrix, $x_2 = x_1^*$, $x_4 = x_3^*$, $x_6 = x_5^*$. The corresponding eigenvalue for $x_1$ is $\lambda_1 = \exp(i\mu_1)$ and the eigenvalue for $x_2$ is $\lambda_1^* = \exp(i\mu_1)$. In a similar way, $\lambda_2$, $\lambda_2^*$ are the eigenvalues for $x_3$ and $x_4$, and $\lambda_3$, $\lambda_3^*$ are the eigenvalues for $x_5$ and $x_6$. The eigenvectors are then normalized so that

$$\begin{aligned} \tilde{x}_1^* \; S \; x_1 &= 2i \\ \tilde{x}_3^* \; S \; x_3 &= 2i \\ \tilde{x}_5^* \; S \; x_5 &= 2i \end{aligned} \tag{4-1}$$

It will then follow that $\tilde{x}_2^* \; Sx_2 = -2i$, $\tilde{x}_4^* \; Sx_4 = -2i$ and $\tilde{x}_6^* \; Sx_6 = -2i$.

Let us now examine the eigenvectors of the one period transfer matrix in the uncoupled coordinate system, $\hat{P} = \overline{R}\hat{T}\;R$ as given by Eq. (2-14). The eigenvectors of $\hat{P}$ will be denoted by $u_1, u_2, u_3, u_4, u_5, u_6$. If $x_1$ is an eigenvector of $\hat{T}$ with the eigenvalue $x_1$ then one can see that $u_1$ given by

$$\begin{aligned} u_1 &= \overline{R} \; x_1 \\ x_1 &= R \; u_1 \end{aligned} \tag{4-2}$$

is an eigenvector of $\hat{P}$ with the same eigenvalue $\lambda_1$. $u_1$ is related to the linear parameters $\beta_1$, $\alpha_1$, $\psi_1$ by Eq. (3-10)

$$u_1 = \begin{bmatrix} \beta_1^{1/2} \\ \beta_1^{1/2}(-\alpha_1 + i) \\ 0 \\ 0 \\ 0 \\ 0 \end{bmatrix} \exp(i\psi_1) \tag{4-3}$$

$x_1$ will be written as

$$x_1 = \begin{bmatrix} x_1 \\ p_{x1} \\ y_1 \\ p_{y1} \\ z_1 \\ p_{z1} \end{bmatrix} \tag{4-4}$$

Putting this result for $u_1$ into Eq. (4-2) and using the result for $R$ given by Eq. (2-16), one finds

$$\begin{aligned} x_1 &= q_1 \beta_1^{1/2} \exp(i\psi_1) \\ p_{x1} &= q_1 \beta_1^{-1/2}(-\alpha_1 + i) \exp(i\psi_1) \\ p_{x1}/x_1 &= (-\alpha_1 + i)/\beta_1 \end{aligned} \tag{4-5}$$

Solving Eqs. (4-5) for $\beta_1$, $\alpha_1$, $\psi_1$ one finds

$$\psi_1 = ph(x_1)$$
$$1/\beta_1 = Im(p_{x1}/x_1) \qquad (4\text{-}6)$$
$$\alpha_1 = -\beta_1 Re(p_{x1}/x_1)$$

where $Im$ and $Re$ stand for the imaginary and real part, and $ph$ indicates the phase.

Using Eq. (4-6), one can find the linear parameters $\beta_1$, $\alpha_1$, and $\psi_1$ from the eigenvector $x_1$ of $\hat{T}$. In a similar way, one can find $\beta_2$, $\alpha_2$, $\psi_2$ from the eigenvector $x_3$, and $\beta_3$, $\alpha_3$, $\psi_3$ from the eigenvector $x_5$.

One may note that one can also compute $q_1$ from Eq. (4-5)

$$q_1 = |x_1|/\beta_1^{1/2} \qquad (4\text{-}7)$$

with similar results for $q_2$ and $q_3$. A procedure is given below for computing the entire $R$ matrix. Also, the tunes of the three normal modes can be computed from the eigenvalues of the $6 \times 6$ transfer matrix $\hat{T}$, $\lambda_i = \exp(i\mu_i)$ $i = 1, 2, 3$. The $\mu_i$ are the phase shifts for a period, and give the fractional part of the tunes.

Having found the $\beta_i$, $\alpha_i$, $\psi_i$ from the one period transfer matrix, one can write the decoupled eigenvectors $u_1$, $u_2$, $u_3$, $u_4$, $u_5$, $u_6$ as

$$u_1 = \begin{bmatrix} \beta_1^{1/2} \\ \beta_1^{-1/2}(-\alpha_1 + i) \\ 0 \\ 0 \\ 0 \\ 0 \end{bmatrix} \exp(i\psi_1), \quad u_3 = \begin{bmatrix} 0 \\ 0 \\ \beta_2^{1/2} \\ \beta_2^{-1/2}(-\alpha_2 + i) \\ 0 \\ 0 \end{bmatrix} \exp(i\psi_2),$$

$$u_5 = \begin{bmatrix} 0 \\ 0 \\ 0 \\ 0 \\ \beta_3^{1/2} \\ \beta_3^{-1/2}(-\alpha_3 + i) \end{bmatrix} \exp(i\psi_3) \qquad (4\text{-}8)$$

$$u_2 = u_1^*, \qquad u_4 = u_3^*, \qquad u_6 = u_5^*$$

Using the decoupled eigenvector $u_i$, and the eigenvectors of $\hat{T}$, $x_i$ one can find a solution for the decoupling matrix $R$.

Let the $6 \times 6$ matrix $X$, and the $6 \times 6$ matrix $U$ be defined as

$$X = (-2i)^{-1/2} [x_1 \ x_2 \ x_3 \ x_4 \ x_5 \ x_6]$$
$$U = (-2i)^{-1/2} [u_1 \ u_2 \ u_3 \ u_4 \ u_5 \ u_6] \qquad (4\text{-}9)$$

where the $x_i$ and $u_i$ are $6 \times 1$ column vectors.

Since the eigenvectors are related by $x_i = R u_i$ one has the relationship

$$X = R\, U \qquad (4\text{-}10)$$

$U$ can be written in terms of $2 \times 2$ matrices as

$$U = \begin{bmatrix} U_{11} & 0 & 0 \\ 0 & U_{22} & 0 \\ 0 & 0 & U_{33} \end{bmatrix}$$

$$U_{11} = \begin{bmatrix} \beta_1^{1/2} \exp(i\psi_1) & \beta_1^{1/2} \exp(-i\psi_1) \\ \beta_1^{-1/2}(-\alpha_1 + i)\exp(i\psi_1) & \beta_1^{-1/2}(-\alpha_1 - i)\exp(-i\psi_1) \end{bmatrix} (-2i)^{1/2} \qquad (4\text{-}11)$$

with similar expressions for $U_{22}$ and $U_{33}$.

The inverse matrix of $U$ can be written as

$$U^{-1} = \begin{bmatrix} U_{11}^{-1} & 0 & 0 \\ 0 & U_{22}^{-1} & 0 \\ 0 & 0 & U_{33}^{-1} \end{bmatrix}$$

$$U_{11}^{-1} = \begin{bmatrix} \beta_1^{-1/2}(-\alpha_1 - i)\exp(-i\psi_1) & -\beta_1^{1/2}\exp(i\psi_1) \\ -\beta_1^{-1/2}(-\alpha_1 + i)\exp(i\psi_1) & \beta_2^{1/2}\exp(i\psi_1) \end{bmatrix} (-2i)^{-1/2} \qquad (4\text{-}12)$$

It can be shown[4] that both $X$ and $U$ are symplectic matrices. Thus $U^{-1} = \overline{U}$ and one can then find for $R$ from Eq. (4-10)

$$R = X\, \overline{U} \qquad (4\text{-}13)$$

Eq. (4-13) provides an explicit solution for $R$ and justifies the assumptions made in section 2 that $R$ is symplectic, periodic and has the form given by Eq. (2-16). Since $X$ and $U$ are both symplectic, then from Eq. (4-13) $R$ is also symplectic. Since the eigenvector $x_1$ has the form $\exp(i\mu s/L)f(s)$, where $f(s)$ is periodic with similar results for the other 5 eigenvectors, all the phase factors in the product $X\overline{U}$ cancel and $R$ is periodic in $S$.

Equations (4-13) and (4-6) give a way to compute the parameters $\beta_i$, $\alpha_i$, $\psi_i$ and the $R$ matrix from the one period transfer matrix $\hat{T} = T(s + L, s)$. It can also be used for coupled motion in 4-dimensional phase space, and provides an alternative procedure to the one given by Edwards and Teng[1] found by solving Eqs. (2-15). The procedure given in this paper may be preferable for use in a tracking program, as it appears more simple to program.

An interesting relationship that can be found from Eq. (4-13) is the connection between $R(s)$ and $R(s_0)$.

$$R(s) = X(s)\overline{U}(s)$$
$$R(s) = \mathrm{T}(s,s_o)X(s_0)\overline{U}(s_0)\overline{P}(s,s_0) \tag{4-14}$$
$$R(s) = \mathrm{T}(s,s_0)R(s_0)P(s_0,s)$$

Eq. (4-14) that relates $R(s)$ and $R(s_0)$ is similar to the result $\hat{\mathrm{T}}(s) = \mathrm{T}(s,s_0)\hat{\mathrm{T}}(s_0)\mathrm{T}(s_0,s)$.

In section 2, the decoupling matrix was defined as the $6 \times 6$ matrix that diagonalizes $\hat{P}(s)$, when written in terms of $2 \times 2$ matrices, according to Eq. (2-14b). It can be shown that $R(s)$ also diagonalizes $P(s,s_0)$. One can write $P(s,s_0)$ in terms of the eigenvectors of $\hat{P}(s)$ as[3]

$$P(s,s_0) = U(s)\overline{U}(s_0) \tag{4-15}$$

Using the result for the decoupled eigenvectors $u_i$ given by Eq. (4-8), one sees that $P(s,s_0)$ is diagonalized when written in terms of $2 \times 2$ matrices.

## 5. The Three Emittance Invariants

Three emittance invariants will be found for linear coupled motion in 6-dimensional phase space. Expressions will be found for these invariants in terms of $\beta_i$, $\alpha_i$. A simple and direct way to find the emittance invariants is to use the definition of emittance suggested by A. Piwinski[4] for 4-dimensional motion. This is given by

$$\epsilon_1 = |\,\tilde{x}_1\ S\ x\,|^2 \tag{5-1}$$

$x$ is a $6 \times 1$ column vector representing the coordinates $x$, $p_x$, $y$, $p_y$, $z$, $p_z$. $x_1$ is a $6 \times 1$ column vector which is an eigenvector of the one period transfer matrix $\hat{\mathrm{T}}$. $x_1$ is assumed to be normalized so that

$$\tilde{x}_1^*\ S\ x_1 = 2i \tag{5-2}$$

One first notes that $\epsilon_1$ given by Eq. (5-1) is an invariant since $\tilde{x}_1\ S\ x$ is a Lagrange invariant as $x_1$ and $x$ are both solutions of the equations of motion. Eq. (4-1) then represents an invariant which is a quadratic form in $x$, $p_x$, $y$, $p_y$, $z$, $p_z$. This result can be expressed in terms of the linear parameters $\beta_1, \alpha_1$ by evaluating $\epsilon_1$ in the coordinate system of the uncoupled coordinates. Since the uncoupled coordinates, represented by the column vector $u$, is related to $x$ by the symplectic matrix $R$,

$$\epsilon_1 = |\tilde{u}_1 \ S \ u|^2 \tag{5-3}$$

$u_1$ is an eigenvector of the one period matrix $\hat{P}$, and one sees that because of Eq. (2-14),

$$x_1 = R \ u_1 \tag{5-4}$$

one can now use the result for $u$, given by Eq. (3-10) and find that

$$\begin{aligned}
\epsilon_1 &= \frac{1}{\beta_1}\left[(\beta_1 p_u + \alpha_1 u)^2 + u^2\right] \\
\epsilon_1 &= \gamma_1 u^2 + 2\alpha_1 u p_u + \beta_1 p_u^2 \\
\gamma_1 &= (1+\alpha_1)^2/\beta_1
\end{aligned} \tag{5-5}$$

One can define $\epsilon_2$ and $\epsilon_3$ as

$$\begin{aligned}
\epsilon_2 &= |\tilde{x}_3 \ S \ x|^2 \\
\epsilon_3 &= |\tilde{x}_5 \ S \ x|^2
\end{aligned} \tag{5-6}$$

and find

$$\begin{aligned}
\epsilon_2 &= \gamma_2 v^2 + 2\alpha_2 v p_v + \beta_2 p_v^2 \\
\epsilon_3 &= \gamma_3 w^2 + 2\alpha_3 w p_w + \beta_3 p_w^2
\end{aligned} \tag{5-7}$$

Eqs. (5-1) and (5-6) can be used in a tracking progress to compute the emittances $\epsilon_1$, $\epsilon_2$, $\epsilon_3$. The tracking program can compute the one period transfer matrix $\hat{T}$ by multiplying the transfer matrices of each element in a period. One can find the eigenfunctions of $\hat{T}$, $x_1$, $x_2$, $x_3$, $x_4$, $x_5$, $x_6$ and normalize them so that

$$\begin{aligned}
\tilde{x}_1^* \ S \ x_1 &= 2i \\
\tilde{x}_3^* \ S \ x_3 &= 2i \\
\tilde{x}_5^* \ S \ x_5 &= 2i
\end{aligned} \tag{5-8}$$

For a given set of the coordinates $x$, $p_x$, $y$, $p_y$, $z$, $p_z$, then $\epsilon_1$, $\epsilon_2$, $\epsilon_3$ can be computed using Eqs. (5-1) and (5-6).

One can also show that the volume in phase space enclosed by these three emittances is given by

$$\int dx \ dp_x dy dp_y dz dp_z = \pi^3 \epsilon_1 \epsilon_2 \epsilon_3 \tag{5-9}$$

The volume in phase space can be computed by going to the coordinate system of the uncoupled coordinates $u$, $p_u$, $v$, $p_v$, $w$, $p_w$ and using $|R| = 1$.

## 6. Summary of Results for Use in a Program to Compute the Linear Parameters and the Decoupling Matrix

The results given above can be used to compute the linear parameters $\beta$, $\alpha$ and $\nu$ and the decoupling matrix $R$ from the one period transfer matrix T. This section will give the steps required to do this using the results given above for particle motion in either 4-dimensional phase space or 6-dimensional phase space.

It is assumed that the one period transfer matrix T is known, having been computed by multiplying the transfer matrices of the elements in one period.

### 6.1. Computing the Linear Parameters $\beta$, $\alpha$, $\nu$

First, compute the eigenvectors and the corresponding eigenvalues of the one period transfer matrix T. To do this one uses one of the standard routines available for computing the eigenvalues of a real matrix. One such set of routines is the routine SGEEV (single precision) and DGEEV (double precision) available in the IBM Engineering and Scientific Subroutine Library (ESSL).

Let us denote the eigenvectors by $x_i$ and the eigenvalues by $\lambda_i$. The eigenvectors and eigenvalues can be arranged in pairs such that $x_2 = x_1^*$ and $\lambda_2 = \lambda_1^*$, and $x_4 = x_3^*$, $\lambda_4 = \lambda_3^*$, and $x_6 = x_5^*$, $\lambda_6 = \lambda_5^*$. Choose $x_1$ to have the eigenvalue $\exp(i\mu_1)$ and $x_2$ to have the eigenvalue $\exp(-i\mu_1)$ and similarly for $x_3$, $x_4$ and $x_5$, $x_6$.

Normalize $x_1$, $x_3$, $x_5$ so that these 3 eigenvectors satisfy

$$\widetilde{x}_i^*\ S\ x_i = 2i \tag{6.1-1}$$

The linear parameters $\beta_1$, $\alpha_1$ can be computed from the normalized $x_1$ using Eq. (4-6)

$$\begin{aligned}1/\beta_1 &= Im\,(p_{x1}/x_1) \\ \alpha_1 &= -\beta_1 Re\,(p_{x1}/x_1)\end{aligned} \tag{6.1-2}$$

$x_1$ and $p_{x1}$ are the first two elements of the column vector $x_1$. The fractional part of the normal mode tune, $\nu_1$ can be computed from the phase advance in a period, $\mu_1$. In the same way one can compute $\beta_2$, $\alpha_2$, $\nu_2$ from $x_3$, $\mu_3$ and $\beta_3$, $\alpha_3$, $\nu_3$ from $x_5$, $\mu_5$.

One may note that the normal mode tunes can be computed directly, without finding the eigenvectors, using the results of reference 6. This may provide a good check of the program.

The diagonal elements of the decoupling matrix $R$ $q_1$, $q_2$, $q_3$ [see Eq. (2-16)] can be found using Eq. (4-2)

$$q_1 = |x_1|/\beta_1^{1/2}$$

## 6.2. Computing the Decoupling Matrix $R$

To compute the decoupling matrix $R$, from the $X$ matrix defined by Eq. (4-9)

$$X = (-2i)^{-1/2} [x_1 \ x_2 \ x_3 \ x_4 \ x_5 \ x_6] \tag{6.2-1}$$

Form the $\overline{U}$ matrix given by Eq. (4-12)

$$\overline{U} = \begin{bmatrix} \overline{U}_{11} & 0 & 0 \\ 0 & \overline{U}_{22} & 0 \\ 0 & 0 & \overline{U}_{33} \end{bmatrix} \tag{6.2-2}$$

The $\overline{U}_{ij}$ are $2 \times 2$ matrices where

$$\overline{U}_{11} = \begin{bmatrix} \beta_1^{-1/2}(-\alpha_1 - i)\exp(-\psi_1) & -\beta_1^{1/2}\exp(-i\psi_1) \\ -\beta_1^{-1/2}(-\alpha_1 + i)\exp(i\psi_1) & \beta_1^{1/2}\exp(i\psi_1) \end{bmatrix} (-2i)^{-1/2}$$

with similar results for $\overline{U}_{22}$ and $\overline{U}_{33}$.

Having formed the matrices $X$ and $\overline{U}$ one can compute $R$ from

$$R = X \ \overline{U} \tag{6.2-3}$$

One may note that above results can be used for particle motion in 4-dimensional phase, or in 6-dimensional phase space with some minor modifications.

### 6.3. Computing the Emittance Invariants

The emittance invariants for a set of coordinates given by the column vector $x$ can be computed using Eq. (5-1)

$$\epsilon_1 = |\: \tilde{x}_1 \; S \; x |^2 \qquad (6.3\text{-}1)$$

where the column vector $x_1$ is an eigenvector of the one period transfer matrix T. Similar results are found for the other invariants $\epsilon_2$, $\epsilon_3$. For $\epsilon_2$ use the eigenvector $x_3$, and for $\epsilon_3$ use the eigenvector $x_5$.